\documentclass[11pt]{article}
\usepackage{jheppub}

\usepackage{amsmath}
\usepackage{amssymb}
\usepackage{verbatim}

\usepackage[normalem]{ulem}

\usepackage[T1,OT1]{fontenc}

\usepackage{graphicx}

\allowdisplaybreaks[4]

\usepackage[hang]{footmisc}
\renewcommand{\d}{\mathrm{d}}
\newcommand{\e}{\mathrm{e}}
\newcommand{\w}{\wedge}

\newcommand{\nl}{\notag \\ &\quad\,}

\title{A note on O6 intersections in AdS flux vacua}

\author{Daniel Junghans}

\affiliation{Institute for Theoretical Physics, TU Wien, Wiedner Hauptstra\ss e 8-10/136, A-1040 Vienna, Austria}

\emailAdd{daniel.junghans @ tuwien.ac.at}

\notoc

\abstract{
The DGKT-CFI construction of AdS flux vacua in type IIA string theory has interesting features such as classical moduli stabilization and a parametric scale separation between the Hubble scale and the Kaluza-Klein scale.
A possible worry regarding the consistency of these vacua is that pathologies could arise due to intersections of the O6-planes, which are not well understood in the 10D solution.
In this note, we show in explicit examples that such intersections are absent if one compactifies on smooth Calabi-Yau manifolds instead of toroidal orbifolds.
In particular, we show that the blow-up of the $T^6/\mathbb{Z}_3$ orbifold yields a single O6-plane which wraps a smooth submanifold without any \mbox{(self-)}intersections.
On the other hand, blowing up the $T^6/\mathbb{Z}_3^2$ orbifold seems to yield an O6-plane which self-intersects. However, we show that this is due to an inconsistent orientifold involution in the original DGKT model.
Imposing a consistent involution again yields a single smooth O6-plane without self-intersections on the blown-up manifold.
}

\begin{document}

\numberwithin{equation}{section}

\maketitle

\newpage

\section{Introduction}

An interesting question both from a phenomenological and a formal point of view is whether string theory has vacua which are 4-dimensional at low energies, i.e., where the extra dimensions are much smaller than the Hubble length. According to the strong AdS distance conjecture (ADC) proposed in \cite{Lust:2019zwm}, this is impossible for supersymmetric AdS vacua. On the other hand, the DGKT-CFI vacua \cite{DeWolfe:2005uu, Camara:2005dc} are potential counter-examples to this conjecture.
These vacua arise in Calabi-Yau orientifold compactifications of type IIA string theory with background fluxes. In the regime where the 4-form fluxes are large, the Hubble scale and the Kaluza-Klein scale are parametrically separated and thus the strong ADC is violated.
An attractive feature of the construction is furthermore that all moduli can be stabilized classically and string corrections are parametrically controlled.\footnote{See also
\cite{Camara:2005dc, Narayan:2010em, Marchesano:2019hfb, Marchesano:2021ycx, Casas:2022mnz, Marchesano:2022rpr} for studies of non-supersymmetric solutions in this setup, \cite{Tringas:2023vzn, Andriot:2023fss} for analyses of various flux scalings, \cite{Aharony:2008wz, Conlon:2021cjk, Apers:2022zjx, Apers:2022tfm, Quirant:2022fpn, Apers:2022vfp, Plauschinn:2022ztd} for holographic aspects
and \cite{Farakos:2020phe, Cribiori:2021djm, Emelin:2022cac, Carrasco:2023hta} for related setups with similar properties.
A refinement of the ADC consistent with DGKT-CFI was proposed in \cite{Buratti:2020kda} but appears to be violated in other examples \cite{Farakos:2020phe, Apers:2022zjx}.}

Further evidence for the existence of the DGKT-CFI vacua was provided in \cite{Junghans:2020acz, Marchesano:2020qvg}, where it was shown that they can be lifted to proper solutions of the 10D equations of motion including the backreaction of the O6-planes. The solutions obtained in \cite{Junghans:2020acz, Marchesano:2020qvg} are valid sufficiently far away from the O6-planes where the backreaction can be treated as a small perturbation around the smeared solution. At large flux, this is true almost everywhere on the internal manifold. What is so far less well understood is the dynamics very close to the O6-planes where their backreaction is non-linear. One might therefore worry that a pathology hides in this so far unaccessible region of the spacetime. In particular, a common worry is that the O6-planes intersect and that some yet to be discovered issue related to the intersection loci could make the solutions inconsistent.
For example, one might speculate that naively ignored degrees of freedom arise at the intersections which could modify the properties of the solutions.

In this note, we argue that this concern is a red herring since there are DGKT-CFI models \emph{without} O6 intersections.\footnote{Our discussion also applies to non-supersymmetric solutions as in \cite{Marchesano:2019hfb}.} This may be surprising since such intersections are \mbox{ubiquitous} on toroidal orbifolds $T^6/\Gamma$, where from the point of view of the covering torus the O6-planes intersect with their images under the orbifold group $\Gamma$. However, as we will show, this is different if we compactify on smooth Calabi-Yau manifolds.
In particular, we give two explicit examples where blowing up the singularities of a toroidal orbifold removes the O6 intersections.

The two examples we study are the $\mathcal{Z}$ manifold and the $\mathcal{Z}/\mathbb{Z}_3$ manifold, i.e., the Calabi-Yau manifolds obtained by blowing up the orbifold singularities of the $T^6/\mathbb{Z}_3$ and $T^6/\mathbb{Z}_3^2$ orbifolds \cite{Candelas:1985en, Strominger:1985it, Dixon:1985jw, Strominger:1985ku, Blumenhagen:1999ev}.
Unlike for other orbifolds, the local metric near the blown-up singularities is known in these cases
\cite{calabi1979metriques, freedman1980remarks, Strominger:1985it}
so that we can study the fixed locus of the orientifold involution very explicitly in these geometries.
Since the $\mathcal{Z}$ and $\mathcal{Z}/\mathbb{Z}_3$ manifolds do not have complex-structure moduli, there is only one special Lagrangian 3-cycle and therefore only a single O6-plane wrapping this 3-cycle. One may therefore suspect that O6 intersections do not arise on these manifolds.

On the $\mathcal{Z}$ manifold, we indeed find that the orientifold involution yields an O6-plane which wraps a smooth submanifold without any \mbox{(self-)}intersections or other types of singularities. Possible concerns about intersections do therefore not apply to this model. On the other hand, on the $\mathcal{Z}/\mathbb{Z}_3$ manifold, the O6-plane seems to self-intersect if one imposes the orientifold involution of the original DGKT model in \cite{DeWolfe:2005uu}. However, we argue that this orientifold involution is actually incompatible with the discrete symmetries of the $T^6/\mathbb{Z}_3^2$ orbifold and thus inconsistent in this particular model. We further show that an alternative involution related to the original one by a sign flip is consistent and does not yield O6-plane intersections on the blown-up Calabi-Yau.

This work is organized as follows. In Section \ref{sec:orbifold}, we briefly review orientifolds of the $T^6/\mathbb{Z}_3$ and $T^6/\mathbb{Z}_3^2$ orbifolds. In Section \ref{sec:metric}, we discuss the blow-up of the orbifold singularities. In Section \ref{sec:o6}, we show that the O6-plane wraps a smooth submanifold in the case of the $\mathcal{Z}$ manifold but seems to self-intersect in the case of the $\mathcal{Z}/\mathbb{Z}_3$ manifold. We also explain that the self-intersections in the second case are related to an inconsistent orientifold involution. In Section \ref{sec:intersections}, we argue that non-intersecting O6-planes in AdS are inconsistent with the equations of motion for tori and toroidal orbifolds but not for Calabi-Yau manifolds.
We conclude in Section \ref{sec:concl} with a discussion of our results and an outlook on future work.

\section{The $T^6/\mathbb{Z}_3$ and $T^6/\mathbb{Z}_3^2$ orientifolds}
\label{sec:orbifold}

Following \cite{Candelas:1985en, Strominger:1985it, Dixon:1985jw, Strominger:1985ku}, we parametrize the $T^6$ by complex coordinates $Z_m$ ($m=1,2,3$) with periodic identifications
\begin{equation}
Z_m \sim Z_m + 1 \sim Z_m+\alpha, \qquad \alpha= \e^{i\pi/3}. \label{torus}
\end{equation}
The torus has a $\mathbb{Z}_3$ symmetry which acts as
\begin{align}
T: Z_m \to \alpha^2 Z_m. \label{qt}
\end{align}
Modding out the torus by $T$, we obtain the $T^6/{\mathbb{Z}_3}$ orbifold. One verifies using \eqref{torus}, \eqref{qt} that this orbifold has 27 singular points. The local geometry in the vicinity of the orbifold singularities is $\mathbb{C}^3/\mathbb{Z}_3$.
One can verify that the orbifold has a further freely acting $\mathbb{Z}_3$ symmetry
\begin{align}
Q: Z_m \to \alpha^{2m} Z_m + \frac{1+\alpha}{3}. \label{qt2}
\end{align}
Modding out by $Q$ yields the $T^6/{\mathbb{Z}_3^2}$ orbifold with 9 singular points. Since $Q$ has no fixed points, the local geometry near the singularities is still $\mathbb{C}^3/\mathbb{Z}_3$.

\begin{figure}[t]
\centering
\includegraphics[trim = 0mm 130mm 30mm 30mm, clip, width=0.6\textwidth]{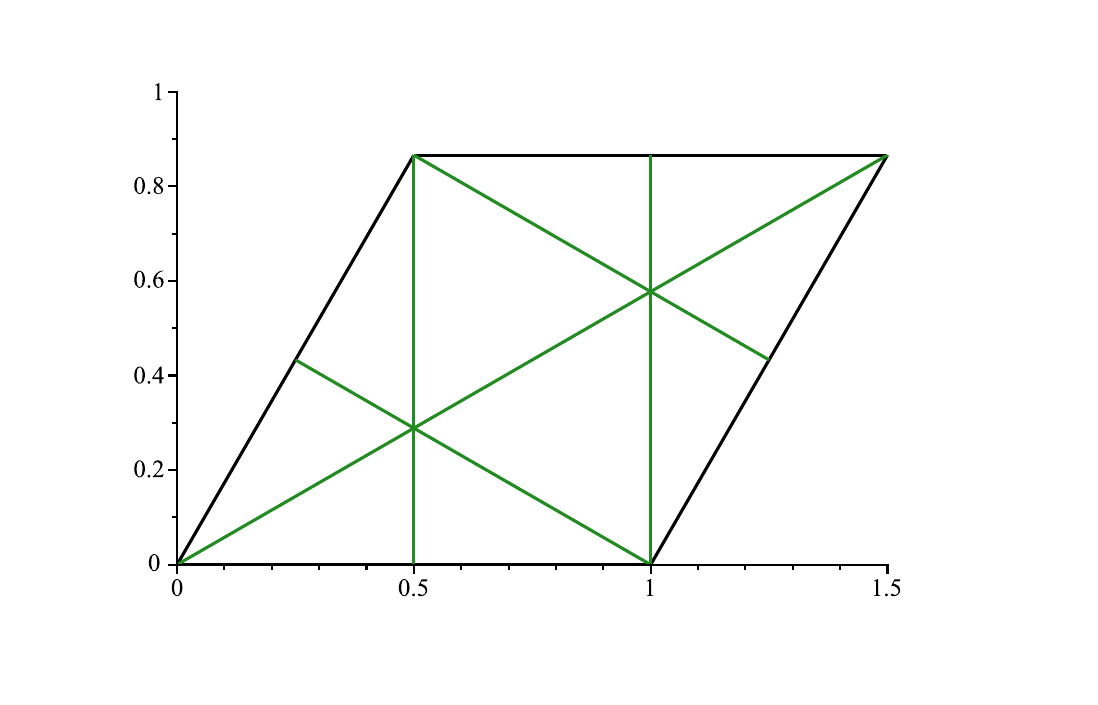}
\put(-260,138){$\scriptstyle{\text{Im}(Z_m)}$}
\put(-85,18){$\scriptstyle{\text{Re}(Z_m)}$}
\caption{Fundamental domain of the 2-tori parametrized by $Z_m$ with O-plane images under $Q$ and $T$ (in green).
\label{fund}}
\end{figure}

We can construct orientifolds of either orbifold using an anti-holomorphic involution \cite{Blumenhagen:1999ev, DeWolfe:2005uu}, which in \cite{DeWolfe:2005uu} was defined as
\begin{equation}
\sigma: Z_m \to -\bar{Z}_m. \label{orientifold}
\end{equation}
Using \eqref{torus} and \eqref{orientifold}, we find that the involution has fixed points at
\begin{equation}
\text{Re}(Z_m) = \frac{\mathbb{Z}}{2}. \label{fp}
\end{equation}
This corresponds to a single O6-plane winding twice around each of the three 2-tori, cf.~Fig.~\ref{fund}. There are also image O-planes, which are generated by $T$ in the case of the $T^6/\mathbb{Z}_3$ orbifold and by $T$ and $Q$ in the case of the $T^6/\mathbb{Z}_3^2$ orbifold.

For the $T^6/\mathbb{Z}_3$ orbifold, acting with $T$ on \eqref{fp} generates two images at $\text{Re}(\alpha^2 Z_m) = \frac{\mathbb{Z}}{2}$ and $\text{Re}(\alpha^4 Z_m) = \frac{\mathbb{Z}}{2}$. Together with \eqref{fp}, we thus have three intersecting O-planes from the point of view of the covering torus, which are mapped to a single O-plane on the orbifold. This O-plane is not smooth at the loci where it intersects with the orbifold singularities. However, we will see in Section \ref{sec:o6} that it is smoothed out by blowing up the orbifold into a Calabi-Yau manifold.

For the $T^6/\mathbb{Z}_3^2$ orbifold, acting with $T$ and $Q$ on \eqref{fp} yields 9 intersecting O-planes on the covering torus, which are again mapped to a single O-plane on the orbifold. As will be discussed in Section \ref{sec:o6}, the O-plane now has self-intersections which are not removed by the blow-up procedure. However, we will see that this happens because the involution \eqref{orientifold} is inconsistent on this particular orbifold and that imposing a consistent involution again yields a smooth O-plane on the blown-up Calabi-Yau.

\section{Metric of the blow-up}
\label{sec:metric}

In order to construct smooth Calabi-Yau manifolds from the orbifolds, we have to cut out each orbifold singularity and glue in a $\mathbb{CP}^{2}$ \cite{Candelas:1985en, Strominger:1985it}.
As explained above, this locally corresponds to blowing up a $\mathbb{C}^3/\mathbb{Z}_3$ singularity. We will focus on the singularity at $Z_m=0$ without loss of generality. The metric of the blow-up of $\mathbb{C}^n/\mathbb{Z}_n$ for general $n$ was discussed in \cite{calabi1979metriques, freedman1980remarks, Strominger:1985it} (see also the later works \cite{Ganor:2002ae, GrootNibbelink:2007lua, Krishnan:2008kv} for discussions of various coordinate choices).
For $n=2$, one obtains the Eguchi-Hanson metric \cite{Eguchi:1978xp}.

The metric for $n=3$ can be written as
\begin{equation}
\d s^2 = g_{m\bar p} \d Y^m \d \bar Y^{\bar p}, \qquad 
g_{m\bar p} = \partial_{m}\partial_{\bar p} K \label{x0}
\end{equation}
in terms of complex coordinates $Y^m$ with $m=1,2,3$ and a K\"ahler potential $K$.
In terms of the $Z_m$ coordinates of the previous section, $K$ is \cite{GrootNibbelink:2007lua}
\begin{equation}
K = \int^S_1 d\tilde S \frac{\left({a^{6}+\tilde S}\right)^{1/3}}{3\tilde S}, \qquad S = \left(\sum_m \bar Z_m Z_m\right)^3, \label{kahler}
\end{equation}
where $a$ is the blow-up parameter.
For $a=0$, this yields $K=S^{1/3} = \sum_m \bar Z_m Z_m$ up to an irrelevant constant and we obtain the flat metric $g_{m\bar p}=\delta_{m\bar p}$ on $\mathbb{C}^3$ and the $\mathbb{C}^3/\mathbb{Z}_3$ orbifold.
For $a\neq 0$, there is a coordinate singularity at
$Z_m=0$, as can be seen
by direct computation of the metric.

To study the origin, we can switch to the local coordinates
\begin{equation}
x=(Z_3)^3, \qquad z_k = \frac{Z_k}{Z_3} \label{coord}
\end{equation}
with $k=1,2$. Note that the transition functions are defined here on the patch $Z_3\neq 0$ but analogous coordinates can be defined with transition functions on the $Z_1\neq 0$ and $Z_2\neq 0$ patches.
Substituting \eqref{coord} into \eqref{kahler} yields
\begin{equation}
S =\bar xx\left(1+\sum_k\bar z_k z_k\right)^3.
\end{equation}
For $a=0$, the orbifold singularity is located at $x=0$. We also note that, according to \eqref{coord}, each point with $x\neq 0$ corresponds to three distinct points on $\mathbb{C}^3$, which indicates that the $x,z_k$ coordinates are coordinates on $\mathbb{C}^3/\mathbb{Z}_3$.
For $a\neq 0$, one can verify that the metric is regular at $x=0$. The orbifold singularity is then replaced by a $\mathbb{CP}^2$ divisor, which is parametrized on a local patch by the $z_k$ coordinates. As usual, three patches are required to cover the full $\mathbb{CP}^2$.

Using
\begin{equation}
x = \frac{\sqrt{r^6-a^6}}{\left(1+ \sum_{k}\bar z_kz_k\right)^{3/2}} \e^{-i\theta}, \label{x3}
\end{equation}
we can write the metric as \cite{Ganor:2002ae}\footnote{Note a few typos in \cite{Ganor:2002ae} compared to our \eqref{x1}, \eqref{x2}.}
\begin{equation}
\d s^2 = \left(1-\frac{a^{6}}{r^{6}}\right)^{-1}\d r^2 + \frac{r^2}{9}\left(1-\frac{a^{6}}{r^{6}}\right) \left(\d \theta - 3A \right)^2 + r^2 \d s_{\mathbb{CP}^2}^2 \label{x1}
\end{equation}
with
\begin{equation}
\d s_{\mathbb{CP}^2}^2 = \sum_{i,j}\frac{\delta_{ij}\left(1+\sum_{k}\bar z_k z_k\right)-\bar z_i z_j}{\left(1+\sum_{k}\bar z_k z_k\right)^2}\d z_i \d \bar z_j, \qquad A = \frac{i}{2} \frac{\sum_{k}\left(z_k \d \bar z_k-\bar z_k\d z_k\right)}{1+\sum_{k} \bar z_k z_k}. \label{x2}
\end{equation}
We thus explicitly see the presence of the $\mathbb{CP}^2$ divisor in the blown-up geometry.
Note that $x=0$ now corresponds to $r=a$, i.e., $r$ parametrizes how far away we are from the (blown-up) singularity.

We finally state a basis of 1-forms in terms of which the metric is diagonal.
In particular, we can define the angular variables \cite{Krishnan:2008kv}\footnote{Our conventions are such that $(r,-\theta, \psi)_\text{here} = (\rho, \psi, \theta)_\text{\cite{Krishnan:2008kv}}$.}
\begin{equation}
x = \sqrt{r^6-a^6} \cos^3\!\sigma\, \e^{-i \theta}, \quad z_1 = \tan \sigma \sin\frac{\psi}{2} \e^{\frac{i}{2}(\phi-\beta)}, \quad z_2 = \tan \sigma \cos\frac{\psi}{2} \e^{-\frac{i}{2}(\phi+\beta)} \label{angul}
\end{equation}
with $\phi-\beta=[0,4\pi)$, $\phi+\beta=[0,4\pi)$, $\psi=[0,\pi]$, $\theta=[0,2\pi)$, $\sigma=[0,\frac{\pi}{2}]$.
The metric thus becomes
\begin{align}
\d s^2 &= \left(1-\frac{a^{6}}{r^{6}}\right)^{-1}\d r^2 + \frac{r^2}{9}\left(1-\frac{a^{6}}{r^{6}}\right) \left(\d \theta + \frac{3}{2}\sin^2\sigma (\d\beta+\cos\psi \d\phi) \right)^2 \nl + r^2 \left( \d \sigma^2+ \frac{1}{4}\sin^2\sigma (\d\psi^2+\sin^2\psi\d\phi^2)+\frac{1}{4}\sin^2\sigma \cos^2\sigma (\d\beta+\cos\psi \d\phi)^2 \right). \label{ang}
\end{align}

\section{O6-planes on the blow-up}
\label{sec:o6}

We now study the O6-plane loci in the blown-up geometries using the metric discussed in the previous section. We first consider the $\mathcal{Z}$ manifold in Section \ref{sec:o61} and then the $\mathcal{Z}/\mathbb{Z}_3$ manifold in Sections \ref{sec:o62} and \ref{sec:o63}.

\subsection{$\mathcal{Z}$ manifold}
\label{sec:o61}

We first recall the orbifold case $a=0$. As explained in Section \ref{sec:orbifold}, the O6-plane and its images under $T$ which pass through the singularity at $Z_m=0$ satisfy
\begin{equation}
\text{O6} :\quad \text{Re}(Z_m)=0, \qquad \text{O6}^{\prime} :\quad\text{Re}(\e^{2i\pi/3} Z_m)=0, \qquad \text{O6}^{\prime\prime} :\quad\text{Re}(\e^{4i\pi/3} Z_m)=0, \label{images}
\end{equation}
respectively. On the covering torus, we thus have that three O6-planes intersect at the orbifold singularity, see Fig.~\ref{fund}. On the orbifold, all three O6-planes are identified and we obtain a single O6-plane. This O6-plane is not smooth at the location where it passes through the singularity.

Using \eqref{coord}, one verifies that \eqref{images} in terms of the $x,z_k$ coordinates becomes
\begin{equation}
\text{O6} :\quad \text{Re}(x) = \text{Im} (z_k)=0. \label{o6}
\end{equation}
Note that \eqref{o6} reproduces all three O-plane images in \eqref{images}.
In the angular coordinates \eqref{angul}, the O6 lies at
\begin{equation}
\text{O6} :\quad \phi+\beta = 2m\pi, \quad \phi-\beta = 2n\pi, \quad \theta = (2l+1)\frac{\pi}{2}, \label{o6p}
\end{equation}
where $m,n,l\in\{0,1\}$. The allowed values are thus $(\phi,\beta)=(0,0),(\pi,\pi),(\pi,-\pi),(2\pi,0)$ and $\theta=\frac{\pi}{2},\frac{3\pi}{2}$. Note that this is still just a single O6-plane which is however described in these coordinates by gluing together 8 pieces.\footnote{This is nothing extraordinary: As a toy example, consider a 1-brane wrapping a longitudinal circle on a 2-sphere. In appropriate coordinates, this is described by two pieces wrapping half-circles at $\varphi=0$ and $\varphi=\pi$ which are glued together at the poles into a single brane.}

We now blow up the orbifold singularity ($a\neq 0$) and study how the O6-plane behaves on the blown-up space. In particular, we would like to know whether the O6-plane has singularities such as cusps or self-intersections.

To this end, we first have to check that \eqref{o6} still determines the O6-plane locus for $a\neq 0$.
Since the O6-plane wraps a special Lagrangian 3-cycle \cite{Becker:1995kb, DeWolfe:2005uu}, we must have $J|_\text{O6}=0$ and $\text{Re}(\Omega)|_\text{O6} = 0$, where $J = i g_{m \bar p} \d Y^m \w \d \bar Y^{\bar p}$ (with $Y^m=x,z_1,z_2$) is the K\"ahler form, $\Omega = f(x,z_1,z_2) \d x\w \d z_1 \w \d z_2$ is the holomorphic 3-form and $f$ is an everywhere non-vanishing holomorphic function. It is straightforward to check by substituting the metric that $J=0$ holds at the locus \eqref{o6} for any $a$. Normalizing $\Omega\w \bar\Omega=-i\text{dvol}_6$, we furthermore have $|f|^2=\frac{1}{8}\text{det} (g_{m\bar p})$, which in the $x,z_k$ coordinates is constant and independent of $a$. Since $f$ is holomorphic, constant $|f|^2$ implies constant $f$. Furthermore, the fact that $|f|^2$ does not depend on $a$ implies that $f$ does not depend on $a$ either, up to an irrelevant constant phase that we may set to zero by convention. It follows that $\Omega$ does not depend on $a$ and thus $\text{Re}(\Omega) = 0$ holds at the locus \eqref{o6} for any $a$. We conclude that the O6-plane on the Calabi-Yau is determined by the same equations \eqref{o6} as in the orbifold case.

\begin{figure}[t]
\centering
\includegraphics[trim = 60mm 100mm 60mm 70mm, clip, width=0.4\textwidth]{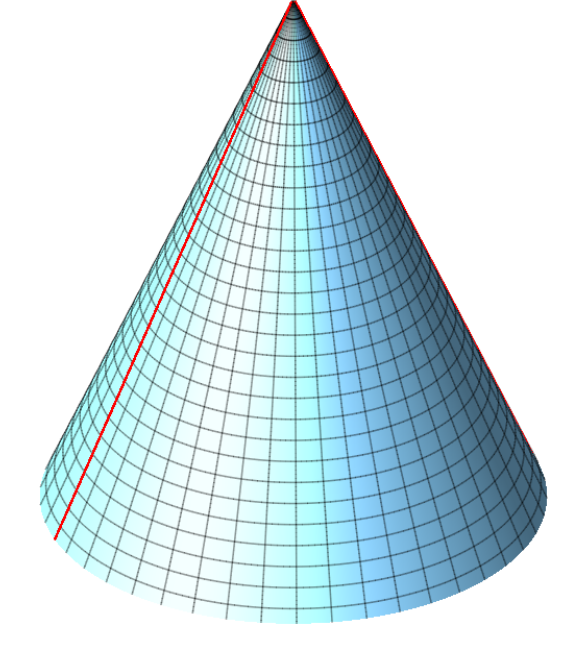}\qquad\qquad \includegraphics[trim = 60mm 100mm 60mm 70mm, clip, width=0.4\textwidth]{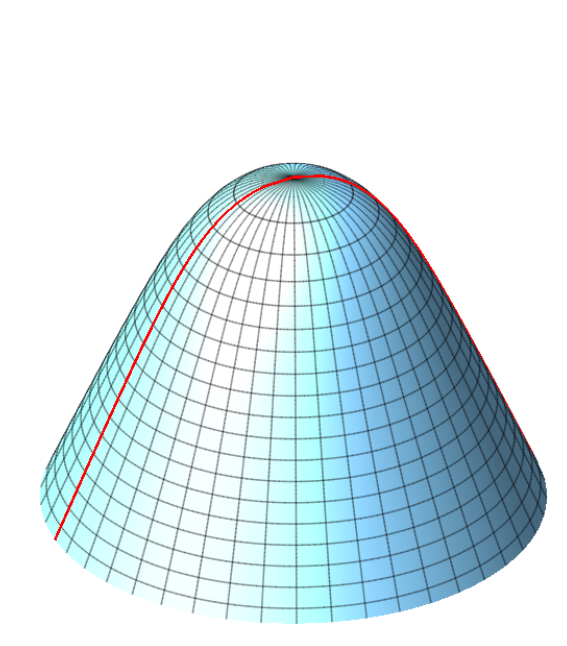}
\put(-205,98){$\scriptstyle{\text{blow-up}}$}
\put(-200,88){$\longrightarrow$}
\caption{2D slice of the local geometry at fixed $z_1,z_2$ for $a=0$ (left) and $a\neq 0$ (right). The O6-plane (red) passes through the orbifold singularity for $a=0$ but is smoothed out when $a$ is finite. \label{geom}}
\end{figure}

We are now ready to study the shape of the O6-plane. A simple first check is to plot a 2D slice of the local geometry at a given point on the $\mathbb{CP}^2$ divisor. As shown in Fig.~\ref{geom}, this suggests that the O6-plane passes smoothly through the blown-up singularity at any such point.

Let us analyze this a bit more carefully. In particular, we also want to understand how the O6-plane behaves along the $\mathbb{CP}^2$ divisor, which is not visible in Fig.~\ref{geom}. For convenience, we will use the notation
\begin{equation}
\text{Re}(z_k) \equiv z_{kr}, \qquad \text{Im}(z_k)\equiv z_{ki}, \qquad \text{Re}(x)\equiv x_r, \qquad \text{Im}(x)\equiv x_i.
\end{equation}
Our interest is in the metric in the vicinity of the blown-up singularity at $x_r=x_i=0$ and the O6-plane at $z_{1i}=z_{2i}=x_r=0$. Expanding the metric in these parameters, it splits at leading order into a sum of the metrics on the tangent and normal spaces of the O6-plane:
\begin{equation}
\d s^2 = \d s^2_\text{tangent} + \d s^2_\text{normal}
\end{equation}
with
\begin{align}
\d s^2_\text{tangent} &= \frac{a^2\left[ (1+z_{2r}^2) \d z_{1r}^2-2 z_{1r}z_{2r} \d z_{1r}\d z_{2r}+(1+z_{1r}^2) \d z_{2r}^2\right]}{(1+z_{1r}^2+z_{2r}^2)^2} + \frac{(1+z_{1r}^2+z_{2r}^2)^3}{9a^4} \d x_{i}^2, \\
\d s^2_\text{normal} &= \frac{a^2\left[ (1+z_{2r}^2) \d z_{1i}^2-2 z_{1r}z_{2r} \d z_{1i}\d z_{2i}+(1+z_{1r}^2) \d z_{2i}^2\right]}{(1+z_{1r}^2+z_{2r}^2)^2} + \frac{(1+z_{1r}^2+z_{2r}^2)^3}{9a^4} \d x_{r}^2.
\end{align}
It is straightforward to verify that $\d s^2_\text{tangent}$ and $\d s^2_\text{normal}$ are smooth and positive definite for all values of $z_{kr}$ except when $z_{1r}$ and/or $z_{2r}$ diverge. In the latter case, we have to switch to a different coordinate patch, where one can then analogously verify that these points are also completely smooth. We conclude that the tangent and normal spaces do not degenerate anywhere, i.e., there are no self-intersections or other singularities.

To study the geometry of the O6-plane, it is convenient to perform the coordinate redefinition
\begin{equation}
z_{1r} = \frac{2w_i}{\bar ww -1}, \qquad z_{2r} = \frac{2w_r}{\bar ww-1}, \qquad x_i=3a^2 y\frac{\left(1-\bar ww\right)^3}{\left(1+\bar ww\right)^3} \label{gsfgsg}
\end{equation}
with $w=w_r+iw_i$. The metric on the tangent space of the O6-plane thus becomes (up to terms $\mathcal{O}(y)$)
\begin{equation}
\d s^2_\text{tangent} = \frac{4a^2}{\left(1+\bar ww\right)^2}\d w \d \bar w + \d y^2.
\end{equation}
The $\mathbb{CP}^2$ divisor replacing the orbifold singularity sits at $y=0$. The first term $\sim \d w\d \bar w$ in the above metric is thus the part along the $\mathbb{CP}^2$, while the second term $\sim \d y^2$ is the part orthogonal to it, corresponding to the red line in Fig.~\ref{geom}. We observe that the first term is the Fubini-Study metric, suggesting that the O6-plane intersects with the $\mathbb{CP}^2$ divisor in a $\mathbb{CP}^1\simeq S^2$. However, on closer inspection, this is not quite true. Indeed, \eqref{gsfgsg} at $y=0$ is invariant under the flip $w\to -1/\bar w$. Hence, antipodal points on the $S^2$ are actually identified, i.e., we have an $S^2/\mathbb{Z}_2 \simeq \mathbb{RP}^2$.
In fact, we could have already guessed this directly from \eqref{o6}, as it is well known that $\mathbb{RP}^2$ can be embedded into $\mathbb{CP}^2$ by restricting the coordinates of the latter to their real parts.
Note that \eqref{gsfgsg} is ill-defined at $|w|=1$ but this is again just a coordinate singularity. We should then simply use a different coordinate patch of $\mathbb{CP}^2$, which yields the same result as on any other patch.

In conclusion, the O6-plane wraps an $\mathbb{RP}^2$ on the blow-up divisor. As stated above, this corresponds to a smooth submanifold of the Calabi-Yau and in particular there are no self-intersections. Note that, although it is well known that $\mathbb{RP}^2$ cannot be embedded into $\mathbb{R}^3$ without self-intersections, this is not true for general manifolds.
Indeed, it is straightforward to check this explicitly here: Let us assume that $\mathbb{RP}^2$ self-intersects on $\mathbb{CP}^2$, i.e., there is a point $w^\prime\neq w$ on $\mathbb{RP}^2$ such that $w$ and $w^\prime$ map to the same point $z_k$ on $\mathbb{CP}^2$. According to \eqref{gsfgsg}, such a point satisfies $\frac{w}{\bar ww -1}=\frac{w^\prime}{\bar w^\prime w^\prime -1}$, which has however no solution except for $w^\prime=w$ and $w^\prime=-1/\bar w$, where the latter is equivalent to $w^\prime=w$ by the antipodal identification. This confirms that $\mathbb{RP}^2$ is smoothly embedded into $\mathbb{CP}^2$ without any self-intersections, in agreement with our earlier discussion of $\d s^2_\text{tangent}$ and $\d s^2_\text{normal}$.

Let us also briefly discuss the angular coordinates \eqref{ang}. The metric near $r=a$ is
\begin{align}
\d s^2_\text{tangent} &= \frac{2a}{3} \d \rho^2 + a^2\d \sigma^2+ \frac{a^2}{4}\sin^2\sigma \d\psi^2, \label{o6a} \\
\d s^2_\text{normal} &= \frac{2a}{3}\rho^2 \left(\d \theta + \frac{3}{2}\sin^2\sigma (\d\beta+\cos\psi \d\phi) \right)^2 \nl + \frac{a^2}{4}\sin^2\sigma \left[ \sin^2\psi\,\d\phi^2+ \cos^2\sigma (\d\beta+\cos\psi \d\phi)^2\right],
\end{align}
where we redefined $r=a+\rho^2$. The (blown-up) singularity thus sits at $\rho=0$.
Recall that $\psi=[0,\pi]$, $\sigma=[0,\frac{\pi}{2}]$ and therefore \eqref{o6a} is the metric of a half-line times $\frac{1}{8}$ of a 2-sphere. This appears to have boundaries at $\psi=0$, $\psi=\pi$, $\sigma=\frac{\pi}{2}$ and $\rho=0$, which cannot be true for the 3-cycle wrapped by the O6-plane. However, according to \eqref{o6p}, the O6-plane in these coordinates is the union of 8 different pieces which all have the same worldvolume metric $\d s^2_\text{tangent}$. Four of these pieces glued together wrap a half-line times half a 2-sphere, and one can verify that they are glued to the remaining four pieces, which also wrap a half-line times half a 2-sphere, in such a way that the boundaries are removed. In particular, this yields an $\mathbb{RP}^2$ at $\rho=0$, consistent with our previous discussion in the $x,z_k$ coordinates.

\subsection{$\mathcal{Z}/\mathbb{Z}_3$ manifold}
\label{sec:o62}

Let us move on to the $\mathcal{Z}/\mathbb{Z}_3$ manifold. We again first discuss the orbifold case $a=0$. As explained in Section \ref{sec:orbifold},
$Q$ and $T$ generate 9 O-plane images at $\text{Re}(Q^l T^p Z_m)=\frac{\mathbb{Z}}{2}$ for $l,p=0,1,2$. Focussing on the local patch near $Z_m=0$ as before, we thus have
\begin{equation}
\text{O6}^{(l,p)} :\quad \text{Re}\left(\e^{2i(lm+p)\pi/3} Z_m\right)=0.
\end{equation}
In terms of the $x,z_k$ coordinates, this becomes
\begin{align}
\text{O6} :\quad &\text{Re} \left( x \right) = \text{Im} \left( z_k \right) = 0, \notag \\ \text{O6}^\prime :\quad &\text{Re} \left( x \right) = \text{Im} \left( z_k \e^{2i\pi k /3} \right) = 0, \notag \\ \text{O6}^{\prime\prime} :\quad &\text{Re} \left( x \right) = \text{Im} \left( z_k \e^{4i\pi k /3} \right) = 0. \label{o6zz}
\end{align}
Recall from Section \ref{sec:metric} that the $x,z_k$ coordinates are coordinates on $\mathbb{C}^3/\mathbb{Z}_3$, which is the local geometry of the $T^6/\mathbb{Z}_3$ orbifold near $Z_m=0$. We thus see that modding out the torus by $T$ maps the 9 images on the torus to 3 images on the orbifold. These 3 images intersect along the line $\text{Re}(x)=z_k=0$ for any $\text{Im}(x)$.
Modding out further by $Q$ maps the 3 images to a single O-plane but crucially this O-plane still self-intersects along $\text{Im}(x)$. The reason is that $Q$ does not have fixed points and therefore the local geometry near any given value of $x,z_k$ (and in particular near the intersection points) is the same as before modding out by $Q$. Indeed, the modding identifies a given point on one of the images with two \emph{distant} points on the other two images but no points are identified within a small neighborhood.
Hence, what globally corresponds to different parts of the same O-plane looks like 3 intersecting O-planes locally.

It is straightforward to see that blowing up the orbifold singularity does not change this conclusion.
Indeed, the O-plane locus is given by \eqref{o6zz} for any $a$, as can be checked by verifying $J|_\text{O6}=\text{Re}(\Omega)|_\text{O6}=0$ analogously to our discussion in Section \ref{sec:o61}. The O-plane therefore self-intersects on the Calabi-Yau at $\text{Re}(x)=z_k=0$ just like it does on the orbifold. As before, modding out by $Q$ does not remove the intersections since this does not change the local geometry.

As a final remark, we point out a subtlety related to the K\"ahler potential. In particular, note that \eqref{kahler} is only invariant under $Q$ for $a=0$ but not for $a\neq 0$. This is seen most easily by expanding $K=K^{(0)}+a^6K^{(1)}+\mathcal{O}(a^{12})$ with
\begin{align}
K^{(0)} &= \bar Z_1Z_1+\bar Z_2Z_2+\bar Z_3Z_3, \qquad K^{(1)} = -\frac{1}{6(\bar Z_1Z_1+\bar Z_2Z_2+\bar Z_3Z_3)^2}.
\end{align}
Using \eqref{qt2}, one verifies that $K^{(0)}$ is invariant under $Q$ up to K\"ahler transformations but $K^{(1)}$ is not.
This may be confusing but can be explained as follows:
While for $a=0$ the K\"ahler potential yields the globally defined metric for the whole torus, for $a\neq 0$ it only yields the (approximate) metric in a local patch near the blow-up of one of the orbifold singularities. As stated before, $Q$ maps points to one another which do not lie within a small neighborhood. This implies in particular that points near one of the blow-up divisors are mapped to points near a different blow-up divisor. Hence, we expect that acting with $Q$ on a local expression for $K$ brings us out of the regime of validity of that local $K$ and instead returns the local $K$ in a different patch. One indeed verifies that this is correct.

\subsection{A closer look at the DGKT orientifold}
\label{sec:o63}

We saw in the previous section that the orientifold involution \eqref{orientifold} leads to the conclusion of a self-intersecting O6-plane on the $\mathcal{Z}/\mathbb{Z}_3$ manifold. We now show that this is actually an artifact of an inconsistent orientifolding.

It is not immediately obvious that self-intersections should be interpreted as a sign of a pathology. Indeed, it is well known that D7-branes in type IIB string theory generically wrap self-intersecting hypersurfaces which are locally Whitney umbrellas \cite{Aluffi:2007sx, Braun:2008ua, Collinucci:2008pf}. These configurations can be understood non-perturbatively using F-theory. On the other hand, there is a simple argument that O-plane intersections on toroidal orbifolds must coincide with an orbifold singularity.\footnote{I am very grateful to Miguel Montero for
making me aware of this argument,
which inspired the analysis that follows in this section.} This is in conflict with our above result (where intersections arise away from the orbifold singularities) and suggests that all intersections should be removed once the singularities are blown up into a smooth Calabi-Yau. The argument goes as follows: Let us denote by $G$ the orbifold action on a point $Z_m$ on the torus and by $\sigma$ the orientifold involution. Now consider an O-plane with the defining equation $Z_m=\sigma(Z_m)$ and one of its images under $G$ satisfying $G(Z_m)=\sigma(Z_m)$. The two O-planes intersect at points where both equations are satisfied and therefore $Z_m=G(Z_m)$ must hold at the intersections. We thus find that the intersections coincide with the fixed points of $G$, proving the claim.

How does this fit together with the result of the previous section?
To resolve this puzzle, recall that the orbifold group of the $T^6/\mathbb{Z}_3^2$ orbifold contains a freely acting $\mathbb{Z}_3$ subgroup, which is generated by
\begin{align}
& Q: Z_m \to \alpha^{2m}Z_m + \frac{1+\alpha}{3}
\end{align}
with $\alpha=\e^{i\pi/3}$ (cf.~Section \ref{sec:orbifold}).
The other $\mathbb{Z}_3$ factor of the orbifold group, which is generated by $T$ as defined in \eqref{qt}, is not relevant for our discussion and will therefore be ignored in the following.
It will further be convenient to view $Q$ as a composition of a rotation $\tilde Q$ and a shift $S$, 
\begin{equation}
\tilde Q: Z_m \to \alpha^{2m}Z_m, \qquad
S: Z_m \to Z_m + \frac{1+\alpha}{3},
\end{equation}
such that $Q(Z_m) = S(\tilde Q(Z_m))$ or, in short, $Q=S \tilde Q$.
Let us also define the orientifold involution
\begin{align}
& \sigma_{\mp}: Z_m \to \mp \bar Z_m, \label{inv2}
\end{align}
where we keep the sign arbitrary for now. Note that the ``minus'' choice is the involution used in the original DGKT model and earlier in this paper (cf.~\eqref{orientifold}).\footnote{Our remarks for the ``minus'' choice in this section also apply to involutions $Z_m \to \e^{i\theta}\bar Z_m$ with $\e^{i\theta}=(\alpha,\alpha^5)$, and our remarks for the ``plus'' choice also apply to involutions with $\e^{i \theta}=(\alpha^2,\alpha^4)$.}

One verifies using \eqref{torus} that
\begin{align}
& S\tilde Q = \tilde Q S, && S \sigma_- = \sigma_- S, && S \sigma_+ = \sigma_+ S^2, && \sigma_-\tilde Q=\tilde Q^2\sigma_-, && \sigma_+\tilde Q=\tilde Q^2\sigma_+ \label{sdjlkgljg}
\end{align}
and
\begin{equation}
S^3=\tilde Q^3=\sigma_\mp^2=1 \label{dslkjsdjlgs}
\end{equation}
up to periodic identifications on the torus. Hence, $S$ and $\tilde Q$ each generate a $\mathbb{Z}_3$ symmetry group of the torus, which we will denote as $\mathbb{Z}_3^{(S)}$ and $\mathbb{Z}_3^{(\tilde Q)}$, and $\sigma_\mp$ generates a $\mathbb{Z}_2$ symmetry group (for either sign).

Ignoring the orientifold $\mathbb{Z}_2$ for the moment, the symmetry group generated by $\tilde Q$ and $S$ is $\mathbb{Z}_3^{(\tilde Q)} \times \mathbb{Z}_3^{(S)}$. We may a priori choose to mod out the torus by any subgroup thereof. In particular, there are four different $\mathbb{Z}_3$ subgroups we can consider: $\mathbb{Z}_3^{(\tilde Q)}$, $\mathbb{Z}_3^{(S)}$ and two ``diagonal'' $\mathbb{Z}_3$'s, which are generated by $S\tilde Q$ and $S^2\tilde Q$, respectively, and will be denoted by $\mathbb{Z}_3^{(S\tilde Q)}$ and $\mathbb{Z}_3^{(S^2\tilde Q)}$. The case of the $T^6/\mathbb{Z}_3^2$ orbifold studied by DGKT and in this paper corresponds to modding out by $\mathbb{Z}_3^{(S\tilde Q)}$.

We now claim that a problem arises if we want to orientifold this orbifold.
Including the orientifold $\mathbb{Z}_2$ symmetry, the relevant generators of the symmetry group are $S$, $\tilde Q$ and $\sigma_\mp$, from which we can build 18 group elements $S^a \tilde Q^b \sigma_{\mp}{}^c$ with $a,b=0,1,2$ and $c=0,1$. All other compositions are equivalent to the mentioned ones upon using \eqref{sdjlkgljg} and \eqref{dslkjsdjlgs}. Importantly, due to the non-trivial commutators \eqref{sdjlkgljg}, the generated symmetry group is not simply $\mathbb{Z}_3^{(\tilde Q)} \times \mathbb{Z}_3^{(S)} \times \mathbb{Z}_2$ but involves semidirect products.
In particular, one obtains
\begin{align}
&& && & \text{``minus'' orientifold: } && (\mathbb{Z}_3^{(\tilde Q)} \rtimes \mathbb{Z}_2) \times \mathbb{Z}_3^{(S)}, && && \label{group} \\
&& && & \text{``plus'' orientifold: } && (\mathbb{Z}_3^{(\tilde Q)} \times \mathbb{Z}_3^{(S)}) \rtimes \mathbb{Z}_2. && && \label{group2}
\end{align}
Due to this non-trivial group structure, some of the orbifoldings mentioned above are only consistent with the ``plus'' orientifold but not with the ``minus'' one.
This is in particular true for the case $\mathbb{Z}_3^{(S\tilde Q)}$ relevant for the orbifold of Section \ref{sec:o62}. Indeed, one verifies using \eqref{sdjlkgljg}
that there is no order-6 subgroup of \eqref{group} generated by $Q=S\tilde Q$ and $\sigma_-$. In other words, the $\mathbb{Z}_2$ generated by $\sigma_-$ is a symmetry of the torus but not of the orbifold. On the other hand, for the ``plus'' orientifold, the elements
\begin{equation}
1,\quad Q,\quad Q^2,\quad \sigma_+,\quad Q \sigma_+,\quad Q^2 \sigma_+
\end{equation}
do form an order-6 subgroup of \eqref{group2}, namely the dihedral group $\mathbb{Z}_3^{(S\tilde Q)} \rtimes \mathbb{Z}_2$. Modding out the torus by $\mathbb{Z}_3^{(S\tilde Q)}$ is therefore consistent with performing the ``plus'' orientifold. Analogous arguments show that modding out by $\mathbb{Z}_3^{(S^2\tilde Q)}$ is also consistent with the ``plus'' orientifold but not with the ``minus'' one, whereas modding out by $\mathbb{Z}_3^{(S)}$ or by $\mathbb{Z}_3^{(\tilde Q)}$ is consistent with both the ``plus'' and the ``minus'' choices. Modding out by the $\mathbb{Z}_3$ generated by $T$ (cf.~\eqref{qt}) is also consistent with both orientifold choices.

\begin{figure}[t]
\centering
\!\!\!\!\!\!\!\!\!\!\!\!\includegraphics[trim = 0mm 0mm 0mm 40mm, clip, width=0.45\textwidth]{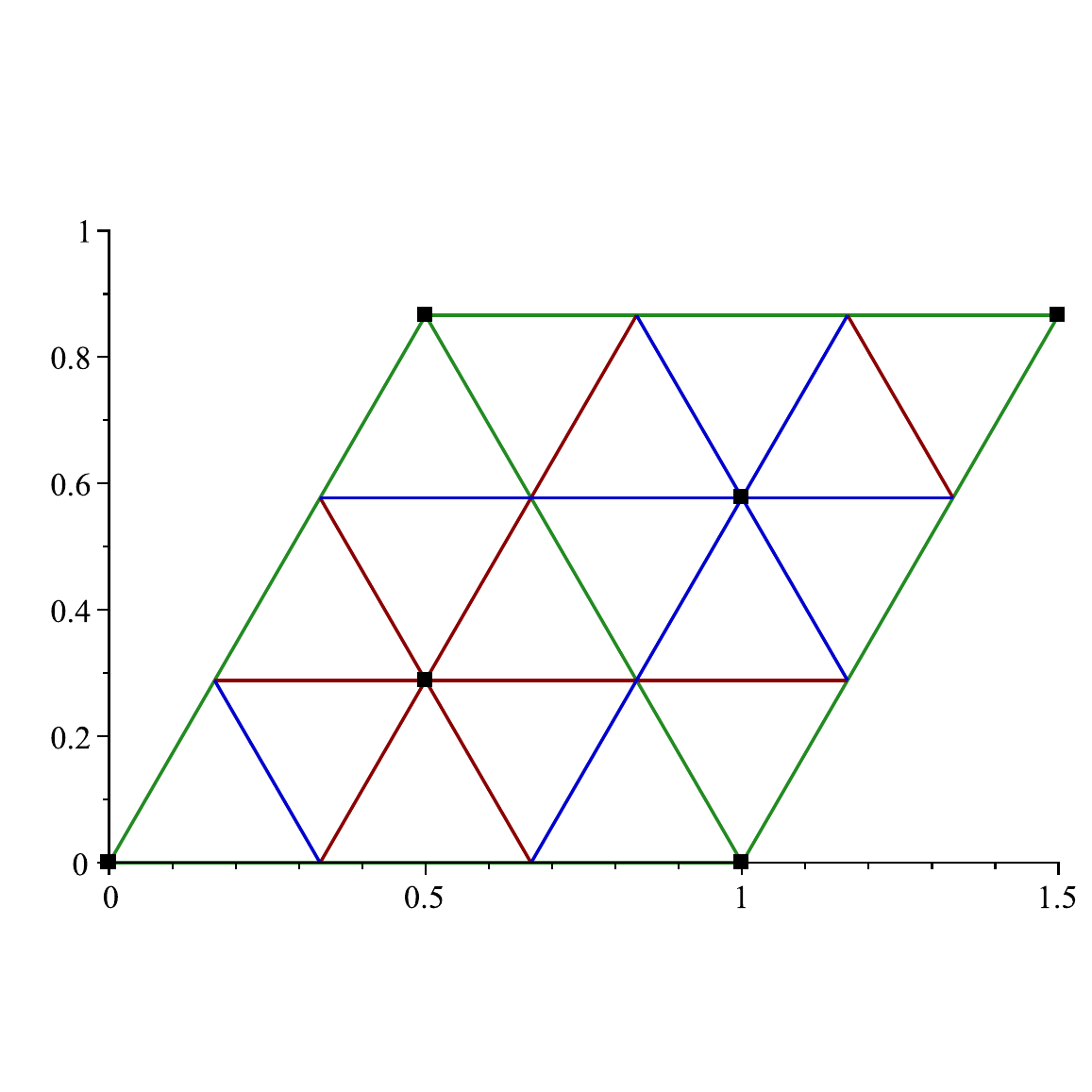}
\put(-217,146){$\scriptstyle{\text{Im}(Z_m)}$}
\put(-42,26){$\scriptstyle{\text{Re}(Z_m)}$}
\\[-0.8em]
\caption{Fundamental domain of the 2-tori parametrized by $Z_m$ with O-plane images under $Q$ and $T$ for the ``plus'' orientifold ($l=0$ in green, $l=1$ in blue and $l=2$ in red). Orbifold singularities are marked as black squares.
\label{fund2}}
\end{figure}

In view of this discussion, it is natural to suspect that the O-plane self-intersections observed in Section \ref{sec:o62} are an artifact of the inconsistent ``minus'' orientifolding and will not be present on the ``plus'' orientifold. One can indeed verify that this is correct.
Repeating the arguments in Sections \ref{sec:orbifold} and \ref{sec:o62} for the ``plus'' orientifold, we get O-plane images at $\text{Im}(Q^l T^p Z_m)=\frac{\sqrt{3}}{2}\mathbb{Z}$ for $l,p=0,1,2$ (see Fig.~\ref{fund2}). This yields
\begin{equation}
\text{O6}^{(l,p)} :\quad \text{Im}\left(\e^{2i(lm+p)\pi/3} Z_m\right)=\frac{3\mathbb{Z}-l}{2\sqrt{3}}. \label{fgkfjlfj}
\end{equation}
Note that, in contrast to Section \ref{sec:o62},
images with different $l$ do not intersect due to the $l$-dependent shift on the right-hand side.
For example, images with $l=0$, $p=0$ satisfy $\text{Im} (Z_1) = 0,\frac{\sqrt{3}}{2}$ on the fundamental domain of the torus, whereas those with $l=1$, $p=2$ satisfy $\text{Im} (Z_1) = \frac{1}{\sqrt{3}}$. Similarly, any other pair of images with different $l$ is parallel and separated on one of the 2-tori parametrized by the $Z_m$ coordinates. Intersections thus only occur between images with the same $l$, and one can verify that they lie on three of the orbifold singularities.

Let us focus again on the local patch near the singularity at $Z_m=0$ as before. In order to pass through this patch, the O-plane images must satisfy \eqref{fgkfjlfj} with a vanishing right-hand side, which implies $l=0$. In terms of the $x,z_k$ coordinates, this becomes
\begin{align}
\text{O6} :\quad &\text{Im} \left( x \right) = \text{Im} \left( z_k \right) = 0.
\end{align}
We thus see that the local orbifold geometry has a single O-plane without self-intersections as in Section \ref{sec:o61}. The analysis of the blow-up also works analogously to that section.

\section{Intersections and equations of motion}
\label{sec:intersections}

In this section, we point out a subtlety related to the consistency of the equations of motion in solutions without O6 intersections. In particular, we show that the equations of motion imply that a DGKT-CFI AdS solution on a torus or toroidal orbifold must have intersecting O6-planes. Crucially, we also explain that this argument does \emph{not} imply O6 intersections on Calabi-Yau manifolds.

To see this, let us assume a (supersymmetric or non-supersymmetric) AdS solution in the general class described in \cite{DeWolfe:2005uu, Marchesano:2019hfb} with a single O6-plane (or several parallel O6-planes).
We consider the smeared solution and follow the conventions stated in \cite{Junghans:2020acz}.\footnote{We further define $|A_p|^2=\frac{1}{p!}A_{MN\cdots P}A^{MN\cdots P}$ and $|A_p|^2_{PQ}=\frac{1}{(p-1)!}A_{MN\cdots P}A^{MN\cdots}{}_Q$ for a $p$-form $A_p=\frac{1}{p!} A_{MN\cdots P} \d x^M \w \d x^N \w \cdots \w \d x^P$ as usual.}
Let us denote the parallel and transverse indices to the O6-plane(s) in the internal dimensions by $m,n,\ldots$ and $a,b,\ldots$, respectively, and assume a basis such that $g_{ma}=0$. We further assume that the internal manifold is Ricci-flat and that $H_3$ has its legs along the transverse directions as in \cite{DeWolfe:2005uu, Marchesano:2019hfb}.

Under these assumptions, the dilaton and Einstein equations yield
\begin{align}
0 &= 2 \mathcal{R}_4 - |H_3|^2 + 2\frac{g_s}{V}, \label{dil} \\
0 &= \frac{1}{4}g_{mn}\mathcal{R}_4 + \frac{1}{2}g_s^2|F_2|^2_{mn} + \frac{1}{2}g_s^2|F_4|^2_{mn}, \label{einst1} \\
0 &= \frac{1}{4}g_{ab}\mathcal{R}_4 + \frac{1}{2}|H_3|_{ab}^2 + \frac{1}{2}g_s^2|F_2|^2_{ab} + \frac{1}{2}g_s^2|F_4|^2_{ab} -g_{ab} \frac{g_s}{V}, \label{einst2}
\end{align}
where
$V$ is the volume of the transverse space (i.e., the 3-cycle on which the O6-plane charge contributes to the tadpole) and $\mathcal{R}_4$ is the AdS scalar curvature.
This implies
\begin{equation}
\mathcal{R}_4 = - \frac{2}{3}g_s^2|F_2|^2_{mn}g^{mn} - \frac{2}{3}g_s^2|F_4|^2_{mn}g^{mn} = - \frac{2}{15}g_s^2|F_2|^2_{ab}g^{ab} - \frac{2}{15}g_s^2|F_4|^2_{ab}g^{ab}. \label{einst4}
\end{equation}
We furthermore know that the components of $F_4$ can at most have 2 transverse indices and that those of $F_2$ have 1 transverse index. One can check that this follows because $H_3$ is transverse by assumption and $H_3 \w \star_6 F_4 = H_3 \w F_2 = \star_6 F_2 F_0 + \star_6 F_4 \w F_2 = 0$ according to the smeared equations of motion and Bianchi identities for the RR/NSNS form fields \cite{Junghans:2020acz}. One furthermore verifies that this index structure implies $|F_4|^2_{mn}g^{mn}\ge |F_4|^2_{ab}g^{ab}$ and $|F_2|^2_{mn}g^{mn}= |F_2|^2_{ab}g^{ab}$.
However, this is inconsistent with \eqref{einst4} unless
\begin{equation}
\mathcal{R}_4=|F_2|^2=|F_4|^2 = 0.
\end{equation}
We thus see that there are no AdS vacua under the above assumptions.

One might be tempted to conclude from this that we always need several intersecting O6-planes in order to solve the equations of motion for AdS.
Indeed, this is true if one assumes a compactification on a flat torus (or an orbifold thereof). The torus is just $\mathbb{R}^6$ with periodic identifications, and the O6-plane extends along an $\mathbb{R}^3$ subspace of it (assuming it is determined by an involution as in \eqref{inv2}).
In any given vacuum solution, we can therefore choose coordinates $x^m$, $x^a$ with $g_{ma}=0$ such that the $x^m$ coordinates are parallel to the O6-plane and the $x^a$ ones are transverse to it. As shown above, the equations of motion then lead to a contradiction. It is therefore impossible to have a DGKT-CFI model on a torus with a single O6-plane (or several parallel ones).

However, on a general Calabi-Yau manifold, a single O6-plane can be consistent with the equations of motion.
The crucial point is that the above argument assumes that the internal coordinates can be split globally into parallel and transverse ones (labelled by $m,n,\ldots$ and $a,b,\ldots$, respectively). Otherwise, we could not have concluded that the smeared O6-plane and $H_3$ flux contribute to the Einstein equations as they do in \eqref{einst1}, \eqref{einst2}. However, such a global split is only possible in simple cases such as tori or products of 3-manifolds. On the other hand, on a Calabi-Yau manifold, we expect that we can only split the coordinates into parallel and transverse ones locally. This means in particular that some parts of the O6-plane can appear ``transverse'' from the point of view of our local coordinates and thus contribute differently to the energy-momentum tensor and the Einstein equations than assumed in \eqref{einst1}, \eqref{einst2}. Similar remarks apply to the $H_3$ flux and the current form $j_3$ of the O6-plane, which have several components when written in terms of basis 1-forms \cite{DeWolfe:2005uu}. While this would indicate intersecting sources in a toroidal model, this conclusion does not follow on a general Calabi-Yau manifold.

\section{Conclusions}
\label{sec:concl}

In this note, we argued that O6-planes in AdS flux vacua need not intersect if one compactifies on a Calabi-Yau manifold instead of a toroidal orbifold.
In particular, we showed that the $\mathcal{Z}$ manifold, i.e., the blow-up of the $T^6/\mathbb{Z}_3$ orbifold, does not yield intersecting O6-planes.
By contrast, we observed that the O6-plane on the $\mathcal{Z}/\mathbb{Z}_3$ manifold seems to self-intersect. However, a detailed analysis of the discrete symmetries in this model revealed that the corresponding orientifold projection is inconsistent. We showed that this inconsistency can be cured by flipping the sign of the orientifold projection compared to the one considered in \cite{DeWolfe:2005uu}
and that in this case no O6-plane intersections arise on the $\mathcal{Z}/\mathbb{Z}_3$ manifold.

Our results thus suggest that the common worry about a pathology related to such intersections may be unjustified in the models we studied:
Indeed, from the point of view of a smooth Calabi-Yau with a single non-intersecting O6-plane, there are no obvious special loci on the O6-plane where anything pathological could happen. A natural expectation is therefore that zooming in on any local patch of the backreacted 10D solution near the O6-plane should yield an ordinary O6 singularity as in flat space. If this is correct, the solution in the orbifold limit (and therefore also the solution for the intersecting O6-planes on the covering torus) does not seem to have an obstruction either, as it can be reverse-engineered by first computing the backreaction on the smooth Calabi-Yau and then setting the volumes of the blow-up cycles to zero. In fact, this technique might also be useful more generally to generate intersecting-brane solutions in flat space, which are notoriously hard to find by direct computation.

Nevertheless, we stress that more work is needed to determine whether the DGKT-CFI construction yields consistent string vacua.
In particular, an important goal for future work is to compute the non-linear backreaction in the near-O6 region and confirm or falsify the expectation outlined above.
The full 10D solution is then given by gluing the near-O6 solution to the solution of \cite{Junghans:2020acz, Marchesano:2020qvg}, which is valid sufficiently far away from the O6.\footnote{A recent numerical analysis in \cite{DeLuca:2021mcj} generically found boundary conditions compatible with a partially smeared O4-plane, whereas obtaining an O6-plane was difficult. It would be important to understand whether this is due to a general obstruction or a limitation of the approach.}
Furthermore, as pointed out in \cite{Junghans:2020acz, Cribiori:2021djm}, it is possible that subleading terms in the large-flux expansion break supersymmetry. It would be important to check this as it could lead to instabilities.

Assuming
a globally defined 10D solution including the full backreaction exists in supergravity, the next step is to understand the stringy resolution of the O6 singularity that is expected to arise at the supergravity level. In particular, a direct M-theory lift of the O6-plane in terms of the Atiyah-Hitchin metric \cite{Atiyah:1985dv, Seiberg:1996nz} is obstructed in DGKT by the Romans mass. However, evidence for a consistent M-theory lift of a T-dual setup was recently given in \cite{Cribiori:2021djm}. We hope to elaborate on this encouraging result and the other open issues in future work.

\section*{Acknowledgments}

I would like to thank Miguel Montero for a helpful correspondence.
This research was funded by the Austrian Science Fund (FWF) under project number P 34562-N.

\bibliographystyle{utphys}
\bibliography{groups}

\end{document}